\documentclass[
reprint,
amsmath,amssymb,
pre,
floatfix,
]{revtex4-1}

\usepackage{graphicx}
\usepackage{dcolumn}
\usepackage{bm}

\usepackage[normalem]{ulem} 
\usepackage{titlesec}
\usepackage{dcolumn}
\usepackage{bm}
\usepackage{graphicx}
\usepackage{afterpage}
\usepackage{xcolor}
\usepackage{amsmath}
\usepackage{amssymb}
\usepackage{graphicx}
\usepackage{bm}
\usepackage{epstopdf}
\usepackage{hyperref}
\usepackage{csquotes}
\hypersetup{
	colorlinks=true,
	linkcolor=blue,
	filecolor=magenta,      
	urlcolor=blue,
	citecolor=red,
}

\newcommand{\eps}{\varepsilon}

\newcommand{\Jav}{J_\mathrm{av}}

\begin{document}

\preprint{betacell1}

\title{Hubs, diversity, and synchronization in FitzHugh-Nagumo oscillator networks: Resonance effects and biophysical implications}

\author{Stefano Scialla}
\email[]{stefano.scialla@unicampus.it}
\affiliation{Department of Engineering, Universit\`a Campus Bio-Medico di Roma - Via \'A. del Portillo 21, 00128 Rome, Italy}

\author{Alessandro Loppini}
\email[]{a.loppini@unicampus.it}
\affiliation{Department of Engineering, Universit\`a Campus Bio-Medico di Roma - Via \'A. del Portillo 21, 00128 Rome, Italy}

\author{Marco Patriarca}
\email[]{marco.patriarca@kbfi.ee}
\affiliation{National Institute of Chemical Physics and Biophysics - R{\"a}vala 10, Tallinn 15042, Estonia}

\author{Els Heinsalu}
\email[]{els.heinsalu@kbfi.ee}
\affiliation{National Institute of Chemical Physics and Biophysics - R{\"a}vala 10, Tallinn 15042, Estonia}

\vspace{0.5cm}

\date{\today}

\begin{abstract}
Using the  FitzHugh-Nagumo  equations to represent the oscillatory electrical behavior of $\beta$-cells, we develop a coupled oscillator network model with cubic lattice topology, showing that the emergence of pacemakers or hubs in the system can be viewed as a natural consequence of oscillator population diversity. 
The optimal hub to non hub ratio is determined by the position of the diversity-induced resonance maximum for a given set of FitzHugh-Nagumo equation parameters and is predicted by the model to be in a range that is fully consistent with experimental observations. 
The model also suggests that hubs in a $\beta$-cell network should have the ability to ``switch on'' and ``off'' their pacemaker function.
As a consequence, their relative amount in the population can vary in order to ensure an optimal oscillatory performance of the network in response to environmental changes, such as variations of an external stimulus.
\end{abstract}


\maketitle



\section{Introduction}
\label{Sec_Intro}

Pancreatic $\beta$-cells in Langerhans islets are characterized by a remarkable coordination of their periodic electrochemical activity, which is linked to their ability to secrete insulin in a pulsatile manner~\cite{Keizer-1988a,Santos-1991a,Benninger-2008a,Rorsman-2018a}. Pulsatile release is thought to be essential for the efficacy of insulin on its target organs and is disrupted in type 2 diabetes~\cite{Bergsten-2000a,Satin-2015a,IdevallHagren-2020a,Laurenti-2020a}. 
This justifies the vast amount of literature aimed at understanding the mechanism of $\beta$-cell electrical oscillations and their synchronization in Langerhans islets, both from the standpoint of cell biology and in terms of biophysical models describing $\beta$-cell clusters as networks of coupled oscillators~\cite{bertram2007metabolic,Pedersen-2009a,Chew-2009a,Goel-2009a,Bertram-2010a,MeyerHermann-2010a,stovzer2013functional,FelixMartinez-2014a,Loppini-2015a,cherubini2015role,markovivc2015progressive,McKenna-2016a,loppini2017gap,loppini2018gap}.

In recent years an increasing number of studies have focused on elucidating the behavior and function of pacemaker cells, also named “hubs” or “leaders”, i.e. subpopulations of $\beta$-cells showing higher oscillatory activity~\cite{Kolic-2016a,Johnston-2016a,Westacott-2017a,Salem-2019a,Lei-2018a,Loppini-2015a,loppini2018emergent}. 
Due to their ability to respond earlier to changes in glucose concentration in the blood stream, hubs would play a crucial role in determining the dynamics of electrical activity of $\beta$-cell clusters, by initiating and synchronizing coordinated electrical oscillations across an islet. While the presence of pacemaker cells in Langerhans islets has been hypothesized several times~\cite{Gylfe-1991a,Aemmaelae-1991a,Palti-1996a,Squires-2002a,Rocheleau-2004a,Benninger-2014a}, the confirmation of their existence via direct observation has become feasible only in recent years, by leveraging new imaging techniques based on optogenetics and recombinant fluorescent probes~\cite{Kolic-2016a,Johnston-2016a,Westacott-2017a,Salem-2019a}.

In spite of this exciting progress and improved understanding, some key questions remain unanswered, specifically: 
(a) Are hubs a permanently distinct subpopulation of $\beta$-cells, or can different $\beta$-cell subsets turn into hubs or non hubs as a function of time and external factors, such as glucose concentration?  
(b) What are the mechanisms that drive the overall frequency of bursting events, i.e. the global oscillatory behavior of an islet as a whole?
While we do not aim to find a definitive solution to these problems, we will show that studying the fundamental dynamical properties of a 3D system of coupled oscillators, mimicking some key features of the electrical behavior of $\beta$-cells, can provide useful insights to understand the collective cell network behavior and to guide future research.

Individual $\beta$-cells that have been isolated from an islet exhibit a heterogeneous electrical activity, ranging from a quiescent state, where their membrane potential stays constantly polarized, to continuous spiking (repeated action potential firings) or bursting events that occur irregularly as a function of time (discrete groups of repeated firings, followed by a period of quiescence)~\cite{Sherman-1988a,Rorsman-1986a,Jonkers-1999a}. 
In contrast, when the same cells are part of an islet, they show strikingly coordinated and regular bursting oscillations, characterized by a period typically ranging from 2 to 5 minutes~\cite{Smolen-1993a,Zhang-2003a}.  
Such membrane potential oscillations are coherent with cytosolic Ca$^{2+}$ ion level fluctuations and correspond to a pulsatile insulin secretion from $\beta$-cells, which is so important for glucose homeostasis and progressively gets lost in type 2 diabetes~\cite{Goldbeter-1990a,Duefer-2004a,Fridlyand-2010a,IdevallHagren-2020a}.

From a dynamical standpoint, bursting activity can be conceived as periodic oscillations of an excitable dynamical system, triggered by an external force that is strong enough to overcome the excitability threshold. 
In the case of $\beta$-cell islets, this force originates from a series of metabolic processes triggered by glucose in the blood stream, therefore it is a function of glucose concentration.

Because of the above mentioned heterogeneity, $\beta$-cells have been a source of inspiration for modeling studies about the effects of diversity on the synchronization of oscillator networks~\cite{Sherman-1994a,Cartwright-2000a}, which has then become a key research topic in complex systems dynamics. 
Numerous studies have documented the emergence of resonance effects, i.e. the amplification of global network oscillations due to diversity, for both bistable and excitable oscillator networks~\cite{Tessone-2006a,Toral-2009a,Chen-2009a,Wu-2010b,Wu-2010a,Patriarca-2012a,Tessone-2013a,Grace-2014a,Patriarca-2015a,Liang-2020a}. 
This effect has been named diversity-induced resonance~\cite{Tessone-2006a} and constitutes an important phenomenon in the context of the present work, where it will be studied by choosing network configurations, topology and coupling relevant to realistic $\beta$-cell clusters.

Indeed, previous studies about diversity-induced resonance focused on $\beta$-cells have not considered the role of pacemakers or hubs, also due to the fact that their existence in Langerhans islets has been confirmed only relatively recently.
The goal of the present work is to investigate whether the existence and key biophysical properties of hubs can be predicted from the general dynamical properties of a network of coupled oscillators mimicking $\beta$-cell electrical behavior.

The paper is organized as follows.
In Sec.~\ref{sec_FN} we summarize the FitzHugh-Nagumo model. 
In Sec.~\ref{sec_hetmodel} we build a coupled oscillator network model that incorporates heterogeneity and cubic lattice topology.
In Sec.~\ref{sec_DIR} we define a metric for estimating the global network oscillation activity and show the emergence of diversity-induced resonance from our model, upon varying oscillator population heterogeneity. 
In Sec.~\ref{sec_HUB} we demonstrate that the presence of pacemakers or hubs in the network can be viewed as a natural consequence of oscillator diversity optimization. 
We also use the model to estimate the percentage of hubs in a network with topological and oscillatory features similar to those of $\beta$-cell clusters in Langerhans islets. 
In Sec.~\ref{sec_ratio} we show that, with respect to the homogeneous system, diversity allows the network to exhibit a more efficient oscillatory response to a range of external stimulus values.
Finally, in Sec.~\ref{sec_conclusion} we discuss the relevance of our results to the understanding of the collective behavior of $\beta$-cells in Langerhans islets, as well as potential correlations with physiological mechanisms underlying pathological conditions, such as type 2 diabetes. 
We also provide perspective on future extensions of this work, such as its comparison to biophysical models and possible applications to other biological systems.


\section{Model}
\label{sec_model}


\subsection{FitzHugh-Nagumo model} 
\label{sec_FN}

Since our aim is to focus mainly on trends and understanding fundamental mechanisms, we will describe individual oscillators by the FitzHugh-Nagumo model, defined by the following dimensionless equations~\cite{Cartwright-2000a,Fitzhugh-1960a,FitzHugh-1961a,Nagumo-1962a}:
\begin{subequations}
\begin{align}
    &\dot{x} = a \left( x - x^3/3 + y \right) \, ,
	\label{eq_FN1a}
    \\
    &\dot{y} = - \left( x  + by - J \right)/a .
	\label{eq_FN1b}
\end{align}
\end{subequations}
Here $x(t)$ is proportional to the membrane potential and $y(t)$ is a recovery variable. 
The quantity $J$ plays the role of an external stimulus and in physiological terms it is related to the glucose level $G$ in the blood stream through some function, $J = f(G)$.
Parameters $a$ and $b$ are proportional, respectively, to the ratio between inductance and capacitance and to the electrical resistance of the $\beta$-cell membrane~\cite{Cartwright-2000a}. 
As will be shown later, they also determine oscillation period and shape.

The above equations are characterized by an equilibrium point, whose stability is determined by the threshold value $\eps$ of the external stimulus $J$:
\begin{align}
    \eps = \frac{3a^2 - 2a^2b - b^2}{3a^3} \sqrt{a^2 - b} \, .
	\label{eq_eps1}
\end{align}
The equilibrium point is stable when $|J|  > \eps$ and unstable when $|J|  < \eps$.  
This means that, when $|J| < \eps$, the system oscillates, while for $|J| > \eps$, it is either in an excitable state ($J < -\eps$), corresponding to a constant negative value of $x(t)$, or in an ``excitation block'' state ($J > \eps$), corresponding to a constant positive value of $x(t)$~\cite{FitzHugh-1961a,Cartwright-2000a}. 
%
\begin{figure*}[ht!]
	\centering
	\includegraphics[width=16cm]{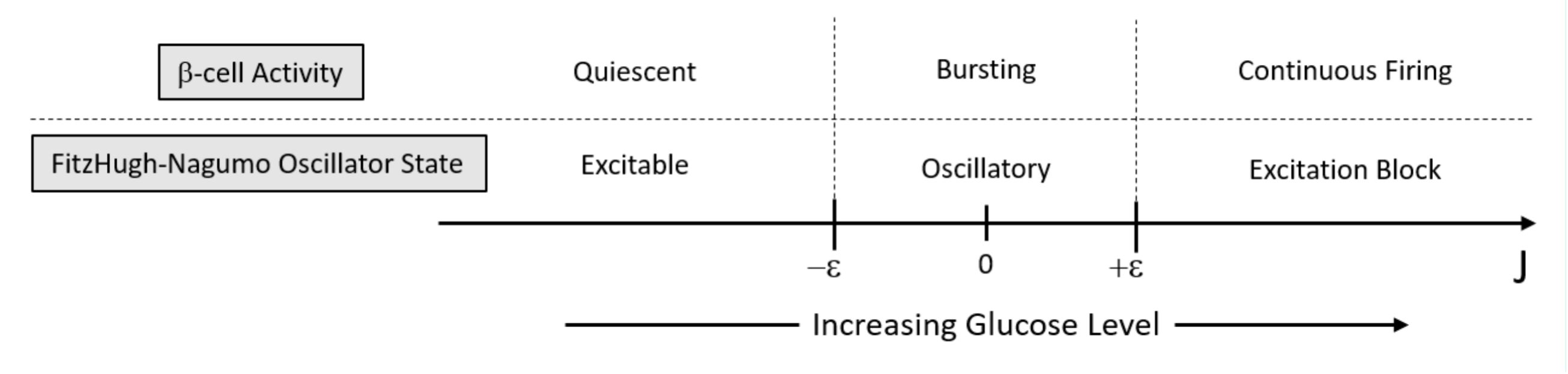}
	\caption{Correspondence between $\beta$-cell activity and FitzHugh-Nagumo oscillator states.
	} 
	\label{figure_1}
\end{figure*}
%
From the standpoint of the electrical behavior of $\beta$-cells, we assume that the interval $|J| < \eps$  corresponds to bursting oscillations, while $J < -\eps$ represents a quiescent polarized state and $J > \eps$ a continuous firing state~\cite{Cartwright-2000a} (see Fig.~\ref{figure_1}).

It may seem strange that we correlate $J$, which can assume both positive and negative values, to glucose level, which is a positive quantity. However, we are not interested in a quantitative correlation between $J$ and glucose level, but want to study trends and mechanisms. Therefore, we just need to keep in mind that $J$ can vary from negative values below $-\eps$, corresponding to  a \textit{low} glucose level; to negative, zero, or positive values in the range $-\eps < J < \eps$, corresponding to \textit{intermediate} glucose levels; and up to positive values above $\eps$, which are representative of \textit{high} glucose levels, see Fig.~\ref{figure_1}. 
Notice that all the values $J > - \epsilon$ correspond to  glucose levels $G_0 \geq G_\textrm{th}>0$, where $G_\textrm{th}$ denotes the activation threshold to induce electrical oscillations in $\beta$-cells.

It is also worth pointing out that $J$ is a constant term in our model equations. This is consistent with most mathematical models on $\beta$-cell electrical activity and is justified by the timescale of bursting, which is much faster than the time required to promote significant glucose variations due to peripheral tissue absorption and hepatic feedback.

\begin{figure}[ht!]
	\centering
	\includegraphics[width=0.45\textwidth]{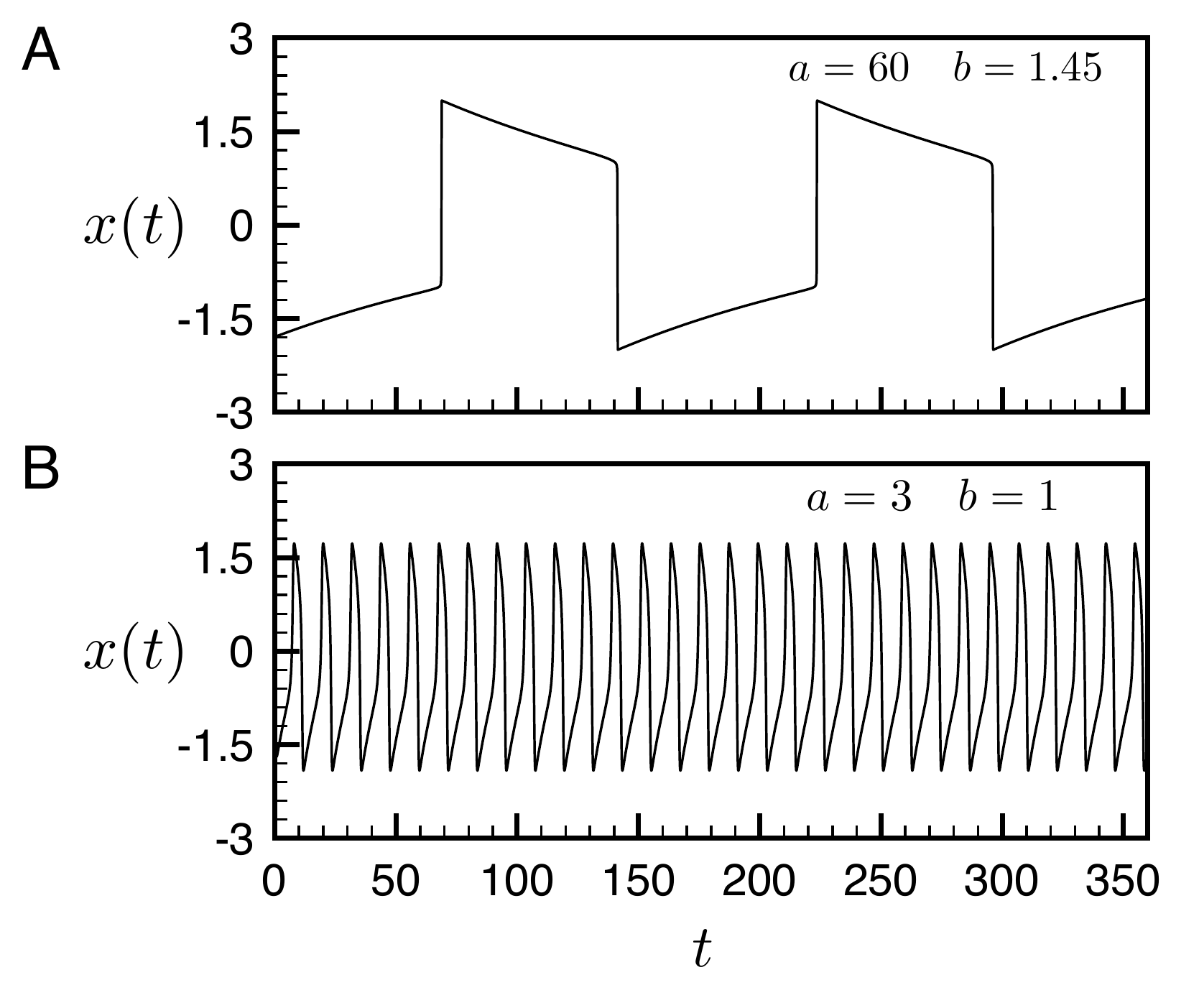}
	\caption{
	Oscillation $x(t)$ of a single FitzHugh-Nagumo element for different values of parameters $a$, $b$ and an external stimulus $J < |\epsilon|$ (see Eq.~\eqref{eq_eps1}), corresponding to the oscillatory regime.
	} 
	\label{figure_2}
\end{figure}

The values of parameters $a$ and $b$ in Eqs.~\eqref{eq_FN1a}-\eqref{eq_FN1b} determine, besides the width of the $|J| < \eps$ interval, the shape of $x(t)$ oscillations. 
Specifically, the oscillation period $T$ is proportional to parameter $a$ (higher values of $a$ correspond to longer oscillation periods), whereas the main effect of parameter $b$ is to modulate the ratio between the time spent by the system at elevated vs. lower $x(t)$.
This is illustrated in Fig.~\ref{figure_2}, showing a comparison between slower (panel A) and faster (panel B) oscillations, corresponding to different combinations of $a$ and $b$ values. 
We will use the combination $a=60$, $b=1.45$ (Fig.~\ref{figure_2}-A) in most of the calculations presented in this work.
If time is expressed in seconds, this combination of values generates a wave with period $T \approx 150$~s and a slightly longer duration of low vs. high $x(t)$ phases, which matches the typical profile of bursting oscillations in $\beta$-cell clusters~\cite{Zhang-2003a}.

It is worth noting that $\beta$-cells have complex dynamical features that are not captured in our approach, i.e. faster action potential spikes superimposed on the slower bursting oscillations, which we reproduce by a FitzHugh-Nagumo description. However, our focus is on the collective dynamics and synchronization of oscillator networks representative of $\beta$-cell clusters, and the role of heterogeneity. In this context, the slower bursting oscillations are more relevant than the action potential spikes, also due to their correlation with pulsatile insulin release, which is critically important from a physiological standpoint.


\subsection{Heterogeneous model}
\label{sec_hetmodel}

In order to describe a $\beta$-cell cluster mimicking a Langerhans islet, we need to build a 3D network of FitzHugh-Nagumo oscillators, which are coupled to their neighbors via coupling factors $C_{ij} (x_j - x_i)$, where $i$ and $j$ are indexes that identify an oscillator $i$ and one of its coupled nearest neighbors $j$. 
We make the simplified assumption that the value of the coupling constant is the same for each oscillator in the network, i.e. it is independent of $i$ and $j$,  $C_{ij} \equiv C$, and that each oscillator is connected to the same number $n$ of neighbors.
Then the corresponding FitzHugh-Nagumo equations for the $i$th oscillator in the network become~\cite{Cartwright-2000a}:
\begin{subequations}
\begin{align}
    &\dot{x}_i = a \left[x_i - x_i^3/3 + y_i + C \sum_{j \in \{n\}_i} (x_j - x_i)\right] \, ,
	\label{eq_FN2a}
    \\
    &\dot{y}_i = - \left( x_i  + by_i - J_i \right)/a ,
	\label{eq_FN2b}
\end{align}
\end{subequations}
where the sum over $j$ in Eq.~\eqref{eq_FN2a} is limited to the set $\{n\}_i$ of the $n$ neighbors coupled to the $i$th oscillator.

In order to introduce diversity in our coupled oscillator network~\cite{Tessone-2006a}, we have assumed in Eq.~\eqref{eq_FN2b} that each oscillator has a different sensitivity to the external stimulus, which is equivalent to associating a different $J_i$ value to each oscillator $i$.
In physiological terms, this can be interpreted as attributing to each $\beta$-cell in an islet a different sensitivity to glucose level, which is a realistic assumption based on available experimental evidence of $\beta$-cell heterogeneity~\cite{Aizawa-2001a,Karaca-2010a,Riz-2014a,Gutierrez-2017a}.

We draw the $J_i$ values from a Gaussian distribution with mean $\Jav$ and standard deviation $\sigma$, which measures the diversity of the oscillator population~\cite{Tessone-2006a}. 
As discussed previously, the mean value of the external stimulus, $\Jav$, is related to glucose level in blood and can therefore be varied in a relatively broad range. 
For simplicity, we initially study the case $\Jav=0$, corresponding to a distribution 
with a certain number of oscillators, depending on the value of $\sigma$, in the oscillatory regime ($|J_i| < \eps$),
and equal numbers of oscillators in the excitable state ($J_i < -\eps$) and in the excitation block state ($J_i > +\eps$).

Using this $J_i$ distribution, we numerically solve the FitzHugh-Nagumo equations for a network of $10^3$ oscillators with cubic lattice topology, where each element is coupled to its six nearest neighbors. 
The $J_i$ values from the Gaussian distribution are randomly assigned to network oscillators throughout the $10 \times 10 \times 10$ cube geometry. 
While the cubic geometry is a simplification, both the total number of oscillators and the number of nearest neighbors per oscillator are consistent with what is known about the structure of Langerhans islets, where each $\beta$-cell is electrically coupled via gap junctions to 6 or 7 neighbor cells on average~\cite{PERSAUD-2014a,Nasteska-2018a}.

We set the coupling constant $C=0.15$ because this value provides an optimal coupling efficiency (lower values cause a steep decrease of global network oscillations, while going higher does not result in a significant increase). 
This is illustrated in more detail in the next section and is a reasonable choice to ensure effective but not unrealistically strong coupling, considering that our goal is to mimic a biological system.

\begin{figure}[ht!]
	\centering
	\includegraphics[width=0.45\textwidth]{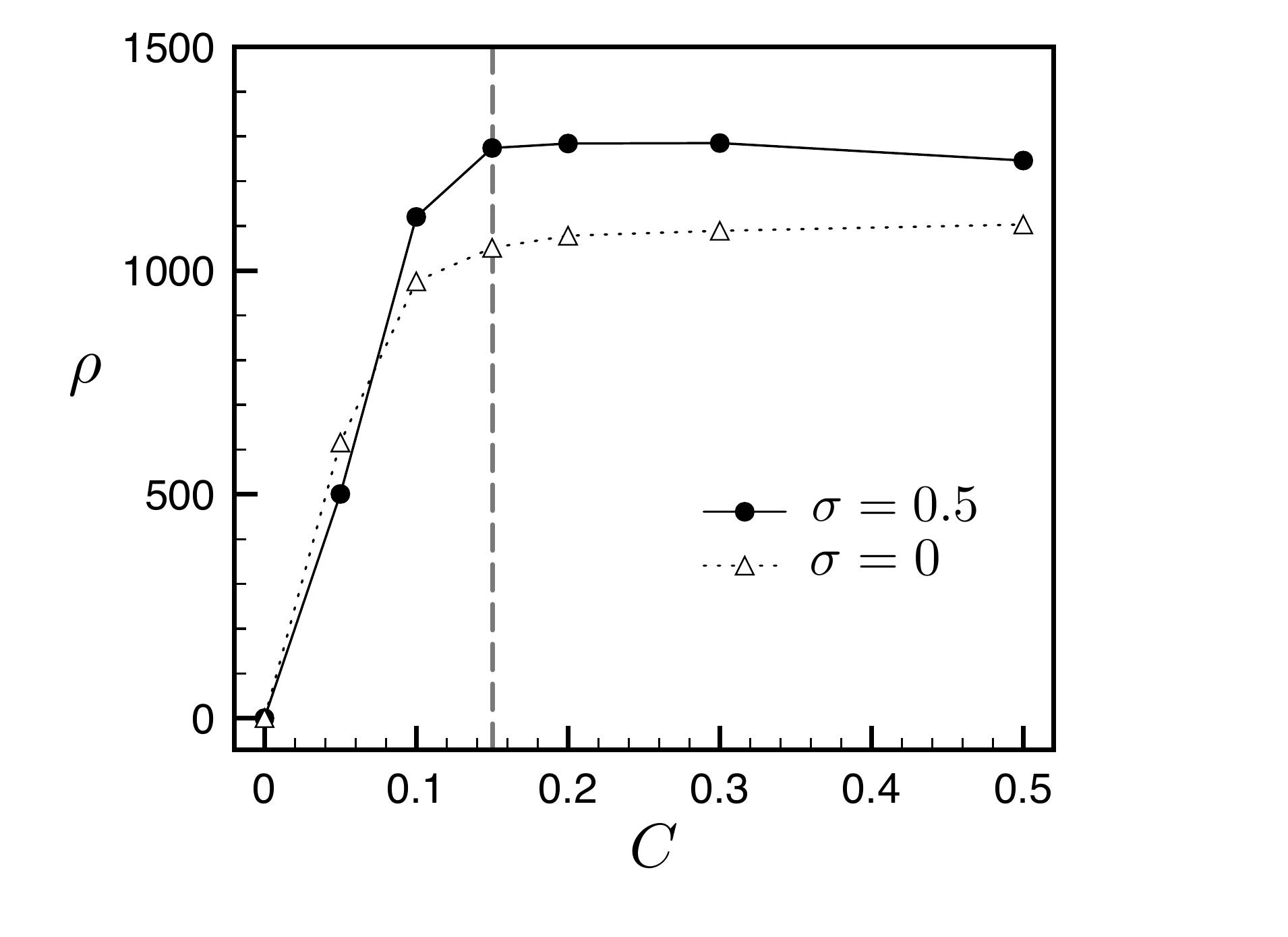}
	\caption{Global oscillatory activity $\rho$ as a function of coupling strength $C$, for different values of population diversity $\sigma$.
	The vertical dashed line at $C=0.15$ corresponds to the coupling strength used in the simulations.}
	\label{figure_3}
\end{figure}

\begin{figure}[ht!]
	\centering
	\includegraphics[width=0.45\textwidth]{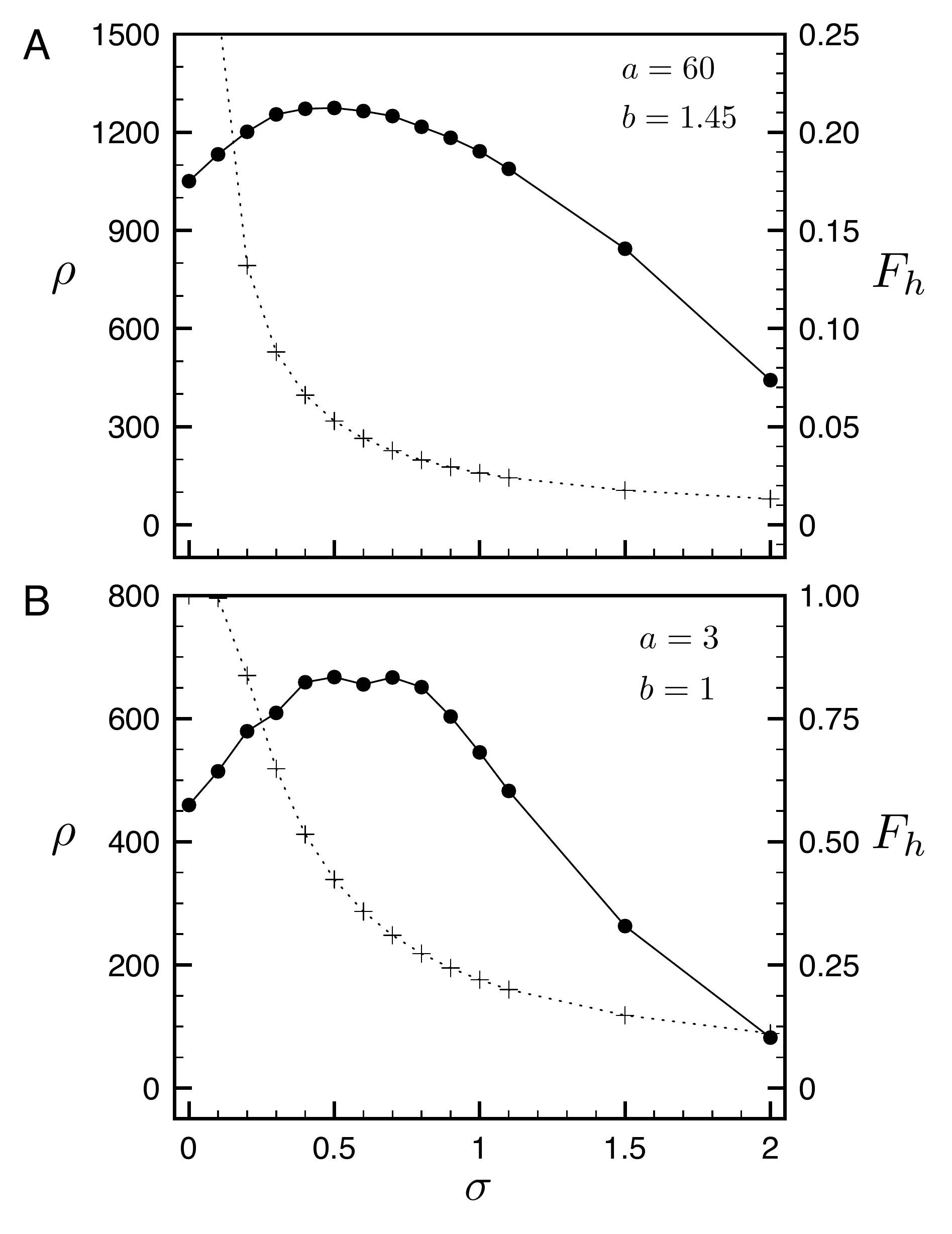}
	\caption{
	Global oscillatory activity $\rho$ (dots, solid curve, left axis) defined in Eq.~\eqref{eq_rho} and fraction of hubs $F_h$ (crosses, dotted curve, right axis) defined in Eq.~\eqref{eq_Fh1} as a function of population diversity $\sigma$, for different values of $a$ and $b$; $\Jav=0$. 
	} 
	\label{figure_4}
\end{figure}


\section{Results}
\label{sec_results}


\subsection{Diversity-induced resonance}
\label{sec_DIR}

After solving the FitzHugh-Nagumo equations \eqref{eq_FN2a}-\eqref{eq_FN2b}, corresponding to the above described topology, we compute the global oscillatory activity of the network~\cite{Cartwright-2000a},
\begin{align}
    \rho = \frac{1}{N} \sqrt{ \frac{1}{t_f} \int_0^{t_{f}} dt \, \left[ X(t) - \bar{X} \right]^2 } .
	\label{eq_rho}
\end{align}
Here $N=10^3$ is the total number of oscillators, $X(t)$ is the sum over all $i$ of the individual $x_i(t)$ functions, and $\bar{X}$ is the mean of $X(t)$ in the time interval $[0,t_{f}]$. 
By its very definition, $\rho$ is the root mean square amplitude over time of the global network oscillation $X(t)$, which has a periodic character.
As a consequence, $\rho$ is substantially independent of $t_{f}$, if $t_{f}$ is sufficiently large. 
We verified that by setting $t_{f}$=300 time units, this condition is satisfied in our calculations.

We simulate numerically the oscillator network for a range of population diversity values $\sigma$, while keeping other parameters constant, i.e. $a=60$, $b=1.45$, and $C=0.15$. 
As mentioned in Sec.~\ref{sec_hetmodel}, this choice of $C$ corresponds to an optimal coupling efficiency, i.e. to the beginning of a plateau when plotting $\rho$ against $C$, as shown in Fig.~\ref{figure_3} for $\sigma=0$ and $\sigma=0.5$.

Using the above parameters, the results for the global oscillatory activity $\rho$ are plotted vs. $\sigma$ in Fig.~\ref{figure_4}-A and show a clear diversity-induced resonance maximum at $\sigma=0.5$. 
This value of $\sigma$ represents the degree of population diversity resulting in the most efficient global network oscillations, due to the interaction between network elements that are individually in an oscillatory regime, i.e. elements for which $|J| < \eps$, and elements that would be, individually, in a non oscillatory regime, due to either quiescence or excitation block state ($|J| > \eps$), but are in fact oscillating due to network coupling and resonance effects. 
Notably, the $\rho$ value corresponding to the diversity-induced resonance maximum is significantly higher than the one achieved with a homogeneous population ($\sigma = 0$) where every element of the network is in the same oscillatory state.


\subsection{Emergence of hubs from diversity optimization}
\label{sec_HUB}

After introducing oscillator diversity via a Gaussian distribution of $J_i$ values and observing the results in terms of global network oscillations, it becomes quite natural to identify network elements corresponding to the interval $|J| < \eps$, which are intrinsically in an oscillatory regime, as pacemakers or ``hubs'' of the system. 
Instead, elements outside the $|J| < \eps$ range are non hubs, which can become active as a consequence of their network interactions and depending on how far their individual values $J_i$ are from the $|J| < \eps$ range.

The hub to non hub ratio corresponding to the diversity-induced resonance maximum represents the most efficient network configuration, because it maximizes global network oscillations. We can estimate this ratio by computing the following normalized Gaussian integral,
\begin{align}
    F_h = \frac{1}{\sqrt{2\pi}\sigma} \int_{-\eps}^{\eps} dJ \, \exp\left[ - \frac{(J - \Jav)^2}{2\sigma^2} \right] ,
	\label{eq_Fh1}
\end{align}
which by definition expresses the fraction of oscillators with $J_i$ values inside the $|J| < \eps$ range, i.e. the fraction of hubs in the population.

The dependence of $F_h$ on $\sigma$ for $a=60$ and $b=1.45$ is shown in Fig.~\ref{figure_4}-A.
The optimal fraction of hubs corresponding to the diversity-induced resonance maximum ($\sigma=0.5$) is $F_h = 0.053$. 
This means a percentage of hubs in the total network population of about 5\%, in good agreement with experimental observations of pacemaker $\beta$-cells in Langerhans islets based on optogenetic methods~\cite{Kolic-2016a,Johnston-2016a,Westacott-2017a,Salem-2019a}, which report this fraction to be 1-10\%.
This prediction of our model is dependent on a specific choice of $a$, $b$ values in Eqs.~\eqref{eq_FN2a}-\eqref{eq_FN2b}, by which we have empirically matched the oscillation period of individual FitzHugh-Nagumo elements with that experimentally observed for $\beta$-cells, as explained in Sec.~\ref{sec_FN}.

The above results show that \emph{in vivo} $\beta$-cell behavior in Langerhans islets, from the standpoint of collective dynamics and network configuration, is consistent with the intrinsic properties of a FitzHugh-Nagumo oscillator network with optimal diversity.
From Fig.~\ref{figure_4}-A one can also see that moving towards higher $\sigma$ values beyond the diversity resonance maximum at $\sigma=0.5$, the slope of $\rho$ becomes progressively more negative and, for $\sigma=2$, where $\rho$ is almost one third of its maximum value, the fraction of hubs, $F_h$, drops to about 1\%. This illustrates the correlation between percentage of hubs and global oscillatory efficiency of the network, and helps understanding what may happen in Langerhans islets, when the optimal hub to non hub ratio is altered by a pathological condition.

For comparison, we repeat the calculations using the values $a=3$ and $b=1$ that correspond to the faster wave in Fig.~\ref{figure_2}-B.  
As shown in Fig.~\ref{figure_4}-B, 
for these values of $a$ and $b$ the global oscillatory activity $\rho(\sigma)$ exhibits a more complex resonance pattern with two maxima, one at $\sigma \approx 0.4$ and the other at $\sigma \approx 0.6$. 
The corresponding $F_h$ values are $F_h=0.52$ and $F_h=0.36$, respectively.

The above comparison indicates that faster global oscillations require a higher relative number of hubs to maintain a good coordination of the oscillator network, which makes sense from both a physical and a physiological standpoint. 
In the case of a slower wave, network elements that are not initially or individually in an oscillatory state have more time to become active and synchronize with hubs via coupling effects, therefore a lower number of hubs is required to obtain efficient global oscillations. 
With a faster wave, synchronization is more challenging and can be achieved only with a sufficiently high percentage of hubs in the network. This difference is deliberately exaggerated in our faster wave example, by choosing very different values of $a$ and $b$ vs. the slower wave example used in our calculations.
However it would be interesting to look for a confirmation of this trend in future experimental work, by comparing the number of detectable hubs in slow vs. fast bursting oscillations of $\beta$-cell clusters.

It is also worth noting that in both combinations of $a$, $b$ parameters we studied, the $\sigma$ value where the diversity-induced resonance maximum occurs is larger than the corresponding $\epsilon$ ($\epsilon \approx 0.033$ for $a=60$, $b=1.45$ and $\epsilon \approx 0.279$ for $a=3$, $b=1$). This may be due to a positive contribution to network resonance from elements that are outside the intrinsic oscillatory range $|J| < \eps$, but not too far away from it. These excitable elements can easily start oscillating and contribute to resonance thanks to coupling. Instead, elements that are far away from the oscillatory range, i.e., at the tails of the distribution, remain quiescent regardless of coupling, therefore are detrimental to global oscillatory efficiency. The best network oscillatory performance is achieved at the diversity-induced resonance maximum, due to an optimal balance of these opposite effects. When $\sigma$ is increased beyond the resonance maximum, the network loses efficiency, because not only the amount of pacemakers and more easily excitable elements decreases but also, at the same time, the amount of the most distant, quiescent elements increases.


\subsection{The optimal hub to non hub ratio maximizes the dynamic range of response to glucose level}
\label{sec_ratio}

We now study what happens when we shift the position of the mean value $\Jav$ of the $J_i$ distribution with respect to the midpoint of the $|J| < \eps$ interval, keeping $\sigma$ constant.
This will give us information about the ability of the oscillator population to cope with a stimulus corresponding to $J$ values that are increasingly distant from the range corresponding to the intrinsic oscillatory regime.

We perform the calculations with $a=60$ and $b=1.45$, corresponding to the reference wave, for three different degrees of diversity: 
$\sigma=0$ (homogeneous system),
$\sigma=0.5$ (the diversity-induced resonance maximum),  
and $\sigma=2.0$ (as an example of large diversity).

\begin{figure}[ht!]
	\centering
	\includegraphics[width=0.45\textwidth]{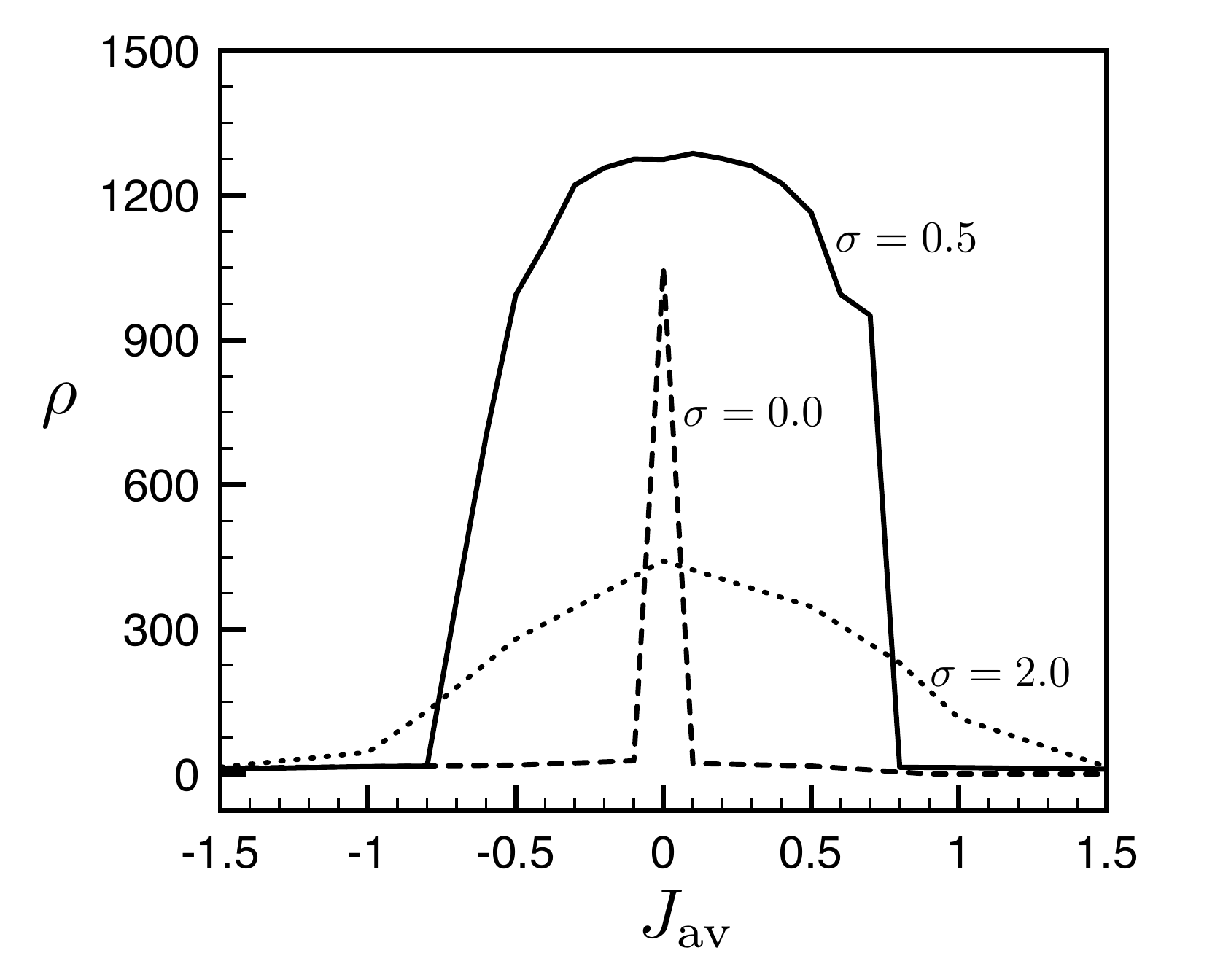}
	\caption{Global oscillatory activity $\rho$, defined in Eq.~\eqref{eq_rho}, as a function of the average value $\Jav$ of the stimulus,
	for different values of population diversity $\sigma$ ($a=60$, $b=1.45$).} 
	\label{figure_5}
\end{figure}

The results reported in Fig.~\ref{figure_5} show that oscillator diversity is able to considerably increase the range of the external stimulus $J$, where the network exhibits efficient global oscillations. 
If all network elements were identical ($\sigma=0$), their global oscillatory activity would be limited to the narrow interval $|J| < \eps \approx 0.033$. 
Instead, oscillator diversity and coupling allow the network to  respond effectively to a much broader range of $J$. 
This range gets broader and broader as $\sigma$ is increased, however, at the same time, increasing $\sigma$ causes a progressively weaker response in terms of global oscillatory efficiency, as shown by the comparison between $\rho$ curves for $\sigma=0.5$ and $\sigma=2.0$.

\begin{figure}[ht!]
	\centering
	\includegraphics[width=0.45\textwidth]{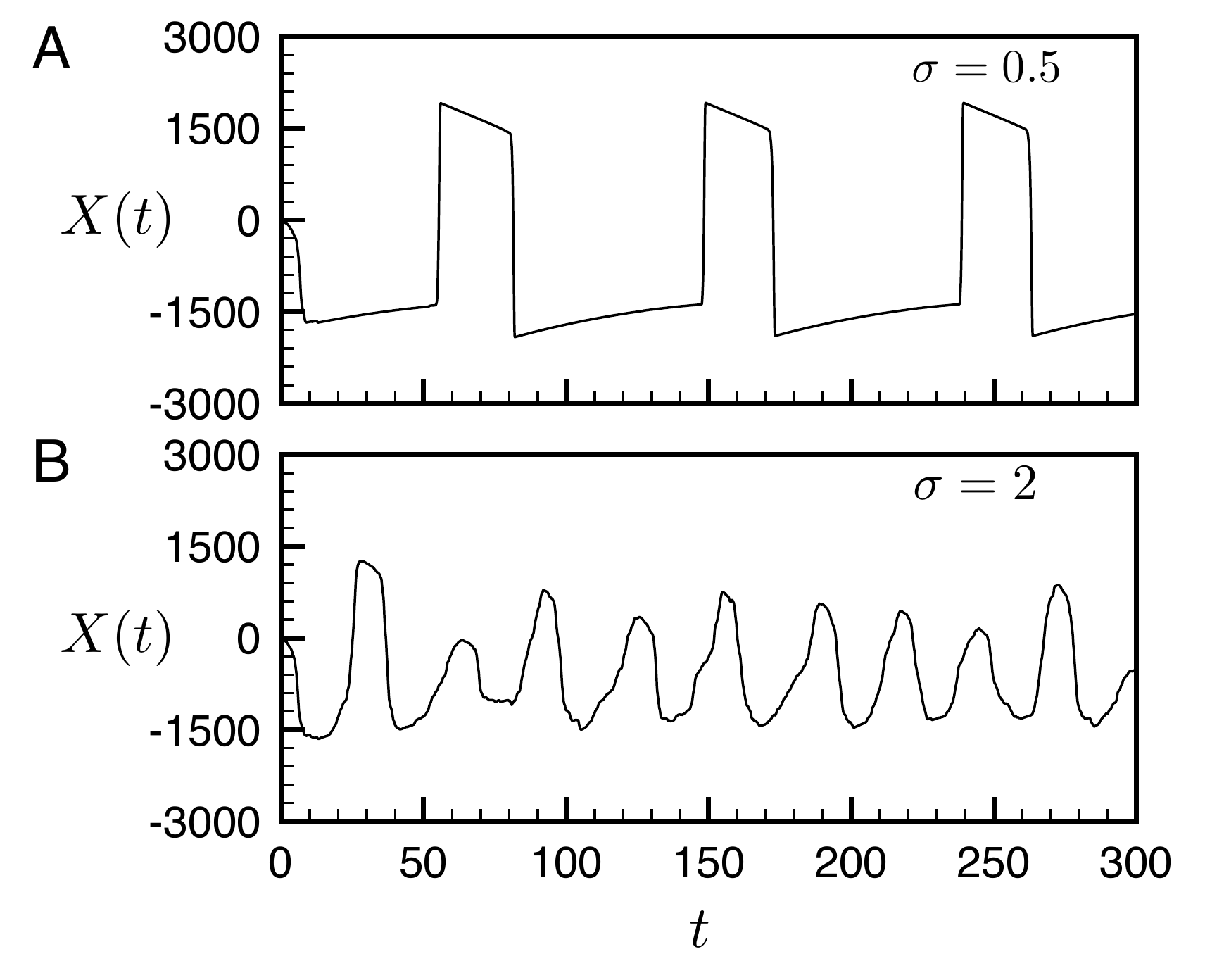}
	\caption{
	Global network oscillation $X(t)$ for different values of population diversity $\sigma$, at the average value $\Jav = 0.5$ of the stimulus ($a=60$, $b=1.45$).
	} 
	\label{figure_6}
\end{figure}

It is also helpful to look at the behavior of  $X(t)$ (the sum of the individual $x_i(t)$) for different values of $\sigma$. 
For instance, for $\Jav=0.5$, the network is in a resonant state and presents global oscillations for both $\sigma=0.5$ and $\sigma=2.0$.
However, a comparison between the corresponding $X(t)$ curves shows a large difference in terms of oscillation amplitude and regularity (Fig.~\ref{figure_6}), which then reflects into very different $\rho$ values for the two parameter sets.
This large difference is a consequence both of a broader $J_i$ distribution, which causes more network elements to have $J_i$ values that are increasingly far away from the oscillatory range $|J| < \eps$, and of a significantly lower number of hubs.

In physiological terms, moving from lower to higher values of $J$ in Fig.~\ref{figure_5} can be considered equivalent to increasing glucose concentration from basal up to elevated levels, as explained in Sec.~\ref{sec_FN}. 
This illustrates that $\beta$-cell diversity can be a mechanism to achieve a much more robust oscillatory behavior of islets in response to varying glucose levels.

It is also interesting to observe that the increase of oscillatory activity from left to right of the $\sigma=0.5$ and $\sigma=2.0$ curves in Fig.~\ref{figure_5} is less steep than the drop on the right side; however, at the same time, the right half of the curve is more extended. 
Again, reading this in physiological terms, we could say that as glucose concentration is gradually increased, the network responds by progressively increasing its oscillatory activity, which is then kept as high as possible for as long as the system is able to cope with the increasing external signal strength. A similar response profile has been predicted also by more complex biophysical models~\cite{Stamper-2019a}, however our approach and analysis helps to clarify and understand the underlying network dynamics.


\section{Conclusions}
\label{sec_conclusion}

Using the FitzHugh-Nagumo equations to represent the electrical behavior of $\beta$-cells, we developed a coupled oscillator network model with cubic lattice topology and showed that the optimization of diversity results in the emergence of pacemakers or hubs, which play a key role in determining the global oscillatory behavior of the network. The optimal hub to non hub ratio predicted by the model is defined by the position of the diversity-induced resonance maximum and depends on oscillation period and shape, which are determined by the FitzHugh-Nagumo equation parameters. 
If we select these parameters in order to match the experimentally measured period of bursting oscillations in $\beta$-cell clusters, we find that the corresponding hub percentage predicted by the model (about 5\%) is in very good agreement with observations of pacemaker $\beta$-cells in Langerhans islets based on optogenetic methods, i.e. $in$ $vivo$ $\beta$-cell behavior in islets is in this respect consistent with the intrinsic oscillatory properties of a heterogeneous, coupled FitzHugh-Nagumo oscillator network embedded in a cubic lattice.

The model also gives an approximate indication of the hub percentage threshold below which the oscillatory performance of a network gets significantly worse, i.e. around 3\%, which may be indicative of the level of $\beta$-cell population alteration corresponding to a pathological condition, such as type 2 diabetes. 
Furthermore, the results obtained suggest the trend that higher bursting oscillation frequencies should correspond to larger hub to non hub ratios, which would be interesting to verify in future experimental work.

We also showed that diversity is a key mechanism to significantly broaden the dynamic range and robustness of the network response to an external stimulus, i.e. glucose concentration in the case of $\beta$-cells. 
This is relevant from a physiological viewpoint and, again, an altered network configuration with suboptimal diversity and hub to non hub ratio will reflect into a compromised oscillatory performance, which in the case of $\beta$-cells translates into an insulin secretion profile that may be insufficient or does not have the required pulsatile characteristics.

Looking back at a key question we asked in the introduction, i.e. whether hubs are a permanently distinct subpopulation of $\beta$-cells, our model suggests that the relative number of hubs in a network can change as a consequence of the external stimulus strength. Therefore, network elements that are non hubs can turn into hubs and vice versa, as the network reconfigures itself in response to an environmental change. Whether hubs are a permanently distinct subpopulation is irrelevant from the standpoint of the dynamical behavior of the oscillator network, however the model suggests that hubs should have the ability to ``turn on'' and ``off'' their pacemaker function in order to ensure optimal network performance in different conditions.

Topics for future extensions of this work include a comparison of the learnings from our approach to biophysical modeling predictions, as well as an in-depth investigation of the combined effects on resonance phenomena of heterogeneity, stochasticity and connectivity, which have been so far partially studied~\cite{gosak2011stochastic}. In addition, we will consider the opportunity to apply our approach or its adaptations to other biological systems beyond $\beta$-cells, e.g. cardiomyocytes and neurons.


\begin{acknowledgments}
    
	A. L. acknowledges the support of Gruppo Nazionale per la Fisica Matematica (GNFM-INdAM). 
	E. H. and M. P. acknowledge support from 
	the Estonian Ministry of Education and Research through Institutional Research Funding IUT39-1, 
	the Estonian Research Council through Grant PUT1356 and PRG1059, 
	and the ERDF (European Development Research Fund) CoE (Center of Excellence) program through Grant TK133. 
	
\end{acknowledgments}



%

\end{document}